# Anomalous Second Harmonic Generation from Atomically Thin MnBi$_2$Te$_4$

Jordan Fonseca$^{1,†}$, Geoffrey M. Diederich$^{1,2,†}$, Dmitry Ovchinnikov$^1$, Jiaqi Cai$^1$, Chong Wang$^3$, Jiaqiang Yan$^{4,5}$, Di Xiao$^{1,3}$, Xiaodong Xu$^{1,3,*}$

$^1$ Department of Physics, University of Washington, Seattle, Washington 98195, United States
$^2$ Intelligence Community Postdoctoral Research Fellowship Program, University of Washington, Seattle, Washington 98195, United States
$^3$ Department of Materials Science and Engineering, University of Washington, Seattle, Washington 98195, United States
$^4$ Materials Science and Technology Division, Oak Ridge National Laboratory, Oak Ridge, Tennessee 37831, United States
$^5$ Department of Materials Science and Engineering, University of Tennessee, Knoxville, Tennessee 37996, United States
$^†$ These authors contributed equally
$^*$ Correspondence to xuxd@uw.edu

**MnBi$_2$Te$_4$ is a van der Waals topological insulator with intrinsic intralayer ferromagnetic exchange and A-type antiferromagnetic (AFM) interlayer coupling. Theoretically, it belongs to a class of structurally centrosymmetric crystals whose layered antiferromagnetic order breaks inversion symmetry for even layer numbers, making optical second harmonic generation (SHG) an ideal probe of the coupling between the crystal and magnetic structures. Here, we perform magnetic field and temperature dependent SHG measurements on MnBi$_2$Te$_4$ flakes ranging from bulk to monolayer thickness. We find that the dominant SHG signal from MnBi$_2$Te$_4$ is unexpectedly unrelated to both magnetic state and layer number. We suggest that surface SHG is the likely source of the observed strong SHG, whose symmetry matches that of the MnBi$_2$Te$_4$-vacuum interface. Our results highlight the importance of considering the surface contribution to inversion symmetry-breaking in van der Waals centrosymmetric magnets.**

MnBi$_2$Te$_4$ has attracted broad interest in recent years due to its potential to host intertwined intrinsic magnetic and topological properties that persist to the atomically thin limit.[1–7] It is a van der Waals crystal in which each stoichiometric septuple layer (SL) can be thought of as a quintuple layer of Bi$_2$Te$_3$ intercalated with an MnTe bilayer (Fig. 1a).[8,9] The Mn spins in each SL couple ferromagnetically to each other with magnetic moments along the easy *c*-axis; however, exchange coupling causes spins in adjacent layers to anti-align, giving rise to A-type AFM ordering below the Néel temperature of $T_N \approx 25\ K$.[1,7] The interplay of thickness, magnetic state, and band topology conspire to produce a rich topological phase diagram. MnBi$_2$Te$_4$ is predicted to host a trivial magnetic insulator in 1 SL flakes, the zero-plateau quantum anomalous hall effect or "axion insulator states" in thin even-layer flakes, and quantum anomalous hall (QAH) states in thin odd-layer flakes.[10]

Despite the excitement precipitated by this theoretical work, robust and consistent experimental evidence for MnBi$_2$Te$_4$'s thickness dependent physics has remained elusive. Shortly after the initial theoretical prediction,[10] the material was synthesized in bulk crystal form and confirmed to be the first AFM topological insulator.[8,9] Subsequent experimental works on atomically thin flakes partially confirmed initial predictions, with the observation of the QAH effect in a 5 SL device[6] and transport signatures of the axion insulator state in 6 SL samples.[11] Nevertheless, reproducing the QAH effect has been difficult. Instead, quantized transport in the field induced ferromagnetic state has been repeatedly observed and become a robust phenomenon in both even- and odd-layer samples. Thus, the nature of MnBi$_2$Te$_4$'s layer-dependent topological and magnetic states remains an open question.

Besides numerous transport studies indicating similar magnetic-state-dependent topological properties of even- and odd-layer number samples,[3,11–17] ARPES measurements provide inconclusive results regarding whether the surface state is gapped at zero magnetic field.[18–20] These experimental incongruities might be related to the impact of atomic defects on magnetism,[21] defect-induced surface reconstruction in few-layer samples,[22] or a possible exfoliation-induced structural or magnetic phase transition,[23–25] all of which could alter the expected layer-dependent magnetism and topology of the studied samples. This situation calls for a non-invasive probe that is sensitive to MnBi$_2$Te$_4$'s layer-dependent structural, magnetic, and surface symmetries.

Optical second harmonic generation (SHG) is an ideal tool to investigate MnBi$_2$Te$_4$'s structural and magnetic properties. Within the 2D materials field, SHG arising from structural inversion symmetry breaking has found utility as a non-invasive means to optically identify crystal axes,[26,27] measure strain,[28] and identify relative twist-angle in layered materials.[29–32] Furthermore, inversion symmetry can be broken by an A-type AFM order (Fig. 1b), leading to so-called *c-type* SHG. It has been employed to explore structural and magnetic properties of a variety of materials including bulk Cr$_2$O$_3$,[33] atomically thin MnPS$_3$,[34] bilayer CrI$_3$,[35] and monolayer MnPSe$_3$.[36] Studies have shown that MnBi$_2$Te$_4$ crystalizes in the centrosymmetric $R\bar{3}m$ ($D_{3d}^5$) space group.[37–39] With A-type AFM magnetic ordering included, this becomes the magnetic space group $R_I$-3c [40] with the crystal and magnetic structure depicted in Figs. 1a, b. In even (odd) layers, strong (zero) *c-type* SHG is predicted,[41] making optical second harmonic generation a feasible probe of both AFM order and the symmetries of the underlying crystal structure with particular sensitivity to the surface.

In this work, we perform polarization resolved SHG and polar reflectance magnetic circular dichroism (RMCD) measurements on MnBi$_2$Te$_4$ flakes ranging in thickness from bulk to monolayer. Contrary to our expectations, we find that the observed SHG is independent of magnetic state, temperature, exfoliation substrate, or crystal thickness. We conclude that any *c-type* SHG arising from the AFM ordering is obscured by a much stronger signal that likely arises from the interface between MnBi$_2$Te$_4$ and vacuum. Our findings are consistent with the expected structural symmetry of MnBi$_2$Te$_4$ and call for additional studies to address the underlying reasons for the absence of observation of *c-type* SHG.

After preparing 1 - 4 SL flakes of MnBi$_2$Te$_4$ (Fig. 1c, Methods), we characterize the layer-dependent magnetic state using RMCD, which is a sensitive optical probe of out-of-plane magnetization for 2D materials.[42] We demonstrate that our layer assignment, magnetization curves, spin-flip coercive fields, and spin-flop/spin-canting fields are consistent with existing literature (Figs. 1d-f, Figure S1a).[4,7] We then employ polarization resolved SHG measurements on the same 1 - 4 SL flakes characterized with RMCD to compare their magnetic state with their SHG response. We measure the co-linear (XX) and cross-linear (XY) SHG response in Faraday geometry, which enables us to maximally probe the elements of the nonlinear susceptibility tensor given the normal incidence restrictions of our measurement apparatus.

At first glance, the SHG response from 2 SL $MnBi_2Te_4$ is consistent with the theoretical prediction (Fig. 2a). A six-fold pattern emerges in both XX and XY channels, which have identical amplitudes and can be fit well by the predicted theoretical model based on AFM time reversal symmetry breaking (Methods).[41] Unexpectedly, the observed pattern remains unchanged under applied magnetic field up to 3 T (Fig. 2b), which is above the 2 SL spin flop field of ~2 T.[4,7] For an SHG signal arising from AFM ordering that breaks inversion symmetry, we expect a change in the underlying magnetic state to manifest as a change to the shape or amplitude of the SHG rotational anisotropy. Experimentally, we do not see this change, which suggests that time reversal symmetry breaking is not the mechanism that explains the observed signal. We more conclusively demonstrate the independence of the measured SHG signal on magnetic ordering with temperature-dependent measurements. As shown in Fig. 2c, the SHG anisotropy and amplitude from the 2 SL remain nearly unchanged up to 150 K (Fig. 2c) which far exceeds the Néel temperature of ~25K.

We observe nearly identical behavior in 1 SL and 3 SL samples. Regardless of magnetic state, all odd-layer flakes of $MnBi_2Te_4$ possess an inversion center located at the middle Mn layer, which should make these layers second-harmonic dark. Surprisingly, we find that the SHG from 1 SL and 3 SL flakes is as strong as that from the 2 SL with identical polarization dependence. As with the 2 SL flake, SHG from 1 SL and 3 SL samples exhibits no magnetic field dependence up to 3 T (Figs. 3a, b) and remains unchanged at temperatures far above the Néel temperature (Figs. 3c, d). Therefore, the observed SHG has no magnetic origin. A 4 SL sample also exhibits nearly identical behavior (Fig. S1). The fluctuations in intensity observed in the temperature dependence data do not reflect a real change in the sample, but rather imperfect experimental conditions across the wide temperature range.

To ensure that our fabrication methods have not introduced any unwanted signal contributions, we explore how the SHG from $MnBi_2Te_4$ depends on flake substrate and crystal thickness. For example, charge transfer between the Au substrate and the atomically thin $MnBi_2Te_4$ flakes we measured could result in a symmetry breaking electric field arising within the crystal.[43] To understand the impact this effect has on our data, we compare the SHG from thin and thick flakes exfoliated onto either Au or $SiO_2$ (Figs. 4a – d). Although there is variation in intensity across the plots shown, no pattern emerged in our measurements that would allow us to identify a consistent trend between flakes exfoliated on gold versus $SiO_2$. Crucially, choice of substrate does not explain the presence of the SHG signal that we observe. Finally, we rule out an exfoliation-induced structural transition by performing SHG measurements at various stages of our flake preparation. We measure SHG from the (0001) surface of bulk $MnBi_2Te_4$ (Fig. 4f) as well as from flakes that have been cleaved onto scotch tape but not deposited onto any substrate (Fig. 4e). We find that the SHG from unexfoliated bulk flakes is typically ~5 times weaker than that measured from exfoliated bulk flakes, but that otherwise the shape and presence of $MnBi_2Te_4$'s rotational anisotropy SHG response is independent of crystal thickness, exfoliation substrate, and exfoliation process. At 800 nm, the nonlinear susceptibility of exfoliated bulk $MnBi_2Te_4$ is ~50 times weaker than the susceptibility of monolayer $WSe_2$ (Fig. S2). While the five-fold change in the SHG intensity across a wide variety of sample thicknesses and preparation methods may be indicative of some symmetry-breaking process that occurs during the exfoliation, it does not explain the presence of SHG in the remainder of our measurements.

Our experimental results conclusively reveal that the dominant second harmonic response of MnBi$_2$Te$_4$ crystals of all thicknesses under 800 nm excitation is unrelated to the material's magnetic properties. Rather, the system exhibits a layer-, magnetic-state-, and temperature-independent response which cannot arise from the centrosymmetric crystal structure of both bulk and monolayer flakes. We note that our experiment does not preclude the existence of the predicted *c-type* SHG, which might be much stronger for mid-IR excitation wavelengths.[41] If there is any c-type SHG, any additional contribution this makes to the dominant *i-type* signal cannot be resolved in our measurements. More specifically, this may be because our excitation energy of ~1.55 eV (800 nm) is much smaller than the spin splitting between the Mn *d*-bands[44] that would give rise to a strong *c-type* SHG signal.

The unexpectedly strong and magnetic-state-independent SHG response of MnBi$_2$Te$_4$ calls for further investigation into the origin of this signal. SHG in centrosymmetric crystals can arise from magnetic dipole or toroidal moments,[45] though we rule out any magnetic contributions to our signal on the grounds that these contributions would depend on applied magnetic field and temperature, contrary to our observations. For a more plausible explanation of the origin of the anomalous SHG signal, we look to MnBi$_2$Te$_4$'s parent materials Bi$_2$X$_3$ (X = Te, Se), both of which exhibit SHG arising from the inversion symmetry broken surface of the crystal where the crystallographic point group symmetry is reduced from $D_{3d}$ to $C_{3v}$,[43,46] as is the case for MnBi$_2$Te$_4$. In Bi$_2$Se$_3$, rotational anisotropy and surface oxidation dependent SHG measurements confirm that the six-fold SHG pattern originates from the anharmonic hyperpolarizability of the Bi-Se and Se-Se bonds at the surface.[43] SHG from MnBi$_2$Te$_4$ also exhibits strong dependence on the surface oxidation (Fig. S3) as well as a rotational anisotropy consistent with the $C_{3v}$ point group. Both of these facts point toward anharmonic polarizability of the in-plane Te-Te bonds at the crystal surface as the main source of the SHG reported here. Because the point group symmetry at an interface is related to the bulk material's point group, our surface SHG measurements corroborate MnBi$_2$Te$_4$'s expected crystal structure and stacking order.

In conclusion, we use RMCD to characterize the layer dependent magnetic ordering of few-layer MnBi$_2$Te$_4$ down to a monolayer. We subsequently employ polarization resolved magneto- and temperature dependent SHG to show that, contrary to expectations, the measured SHG pattern at 800 nm excitation is independent of the material's magnetic ordering and layer thickness. We emphasize that our most important finding is that the SHG generated from widely available Ti:sapphire sources does not probe AFM magnetic order in MnBi$_2$Te$_4$, but rather likely arises from an inevitable structural symmetry breaking at the crystal surface. Our results lay the experimental groundwork for nonlinear optical probe of the magnetic and structure properties in topological antiferromagnet MnBi$_2$Te$_4$.

**Supporting Information:** Experimental methods, 4 SL RMCD & SHG, SHG from MnBi$_2$Te$_4$ compared to WSe$_2$, Effect of surface oxidation on SHG, and laser-induced heating RMCD.

**Acknowledgements:** We thank Kyle Seyler for helpful discussion during the preparation of this manuscript. This work was mainly supported by the Department of Energy, Basic Energy Sciences, Materials Sciences and Engineering Division (DE-SC0012509). Magento-optical measurement is partially supported as part of Programmable Quantum Materials, an Energy Frontier Research Center funded by the U.S. Department of Energy (DOE), Office of Science, Basic Energy Sciences (BES), under award DE-SC0019443. Device fabrication is partially supported by AFOSR FA9550-21-1-0460. JY acknowledges support from the U.S. Department of Energy, Office of Science, Basic Energy Sciences, Materials Sciences and Engineering Division. The authors also acknowledge the use of the facilities and instrumentation supported by NSF MRSEC DMR-1719797. This research was supported by an appointment to the Intelligence Community Postdoctoral Research Fellowship Program at the University of Washington, administered by Oak Ridge Institute for Science and Education through an interagency agreement between the U.S. Department of Energy and the Office of the Director of National Intelligence. XX acknowledges the support from the State of Washington funded Clean Energy Institute.

**Author Contributions:** X.X conceived the experiment. J.F. prepared samples, assisted by D.O. and J.C. G.M.D. and J.F. performed the RMCD and SHG measurements. C.W. and D.X. verified the SHG theoretical calculations. J.Y. synthesized and characterized the bulk crystals. J.F., G.M.D., D.O., and X.X. analyzed the data and wrote the paper, with input from all authors. All authors discussed the results.

**Competing Financial Interests:** The authors declare no competing financial interests.

**Data Availability**: The data that support the findings of this study are available from the corresponding authors upon reasonable request.

**Figures**

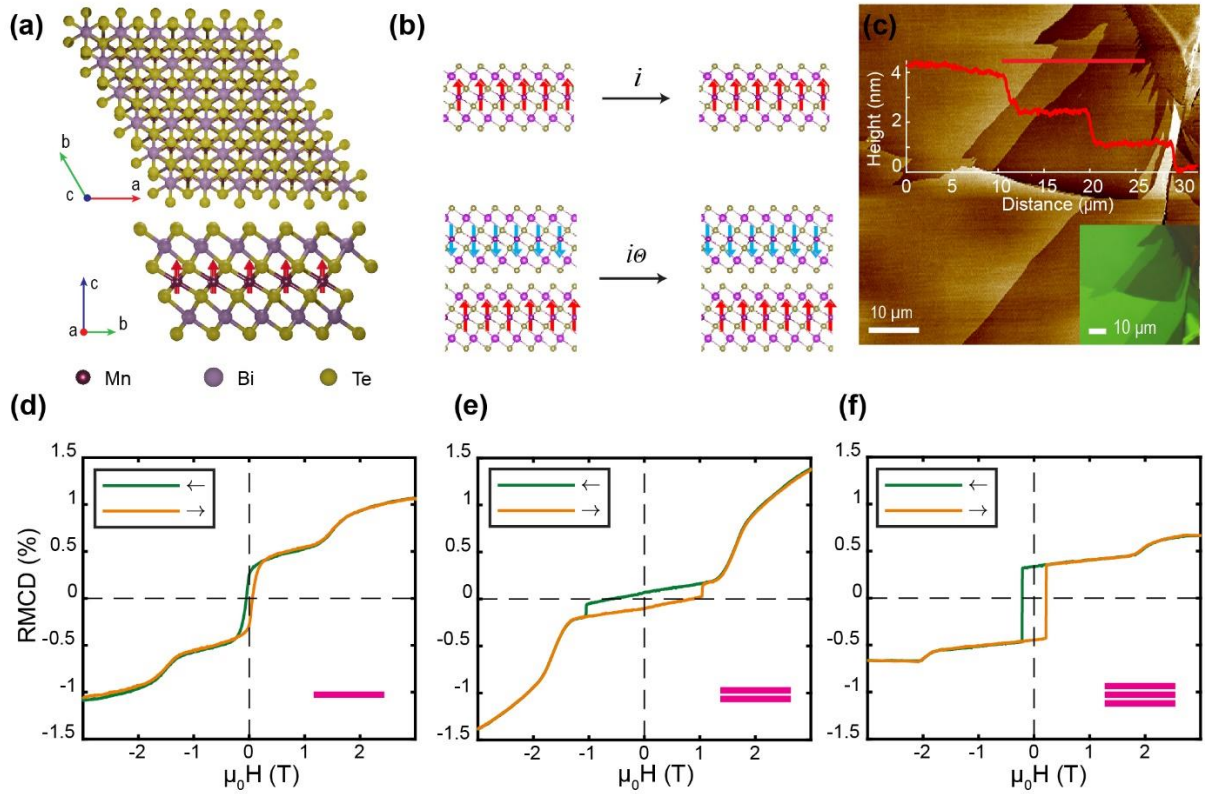

**Figure 1: MnBi$_2$Te$_4$ crystal structure and magnetic ordering.** **(a)** MnBi$_2$Te$_4$ crystal structure. **(b)** Schematic showing how symmetry is preserved (broken) in odd (even) layer crystals. Red/blue arrows indicate the magnetic ground state below the Néel temperature. **(c)** Atomic force microscope image of MnBi$_2$Te$_4$ flakes down to a monolayer with linecut of 1 SL, 2 SL, and 3 SL steps taken along red line. Inset shows optical image of the same region. **(d)-(f)** RMCD sweeps of 1 SL, 2 SL, and 3 SL flakes measured at ~4.5 K. Magenta insets indicate MnBi$_2$Te$_4$ layer number measured in each plot.

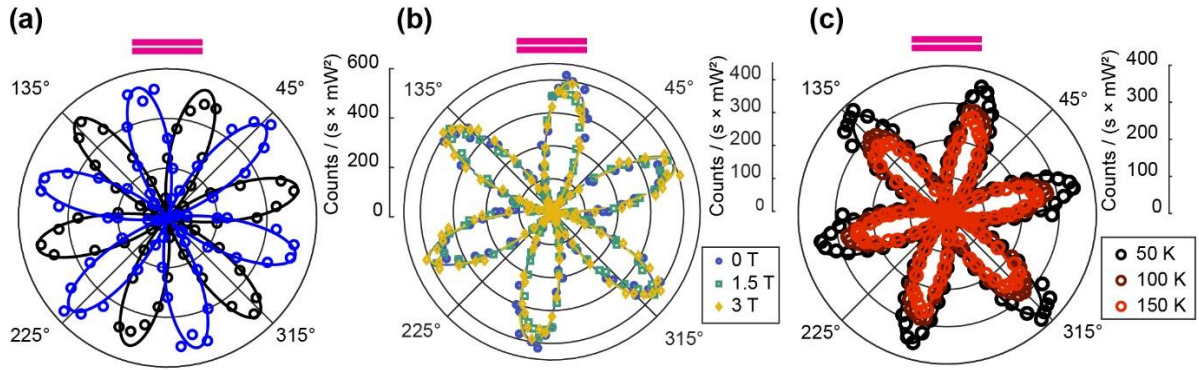

**Figure 2: 2 SL MnBi$_2$Te$_4$ magnetic field and temperature dependent SHG.** **(a)** Polarization-resolved SHG from 2 SL MnBi$_2$Te$_4$ with no applied field below the Néel temperature. Black (blue) dots are co(cross)-linearly polarized experimental data with solid line fits to Eq. 1 (Eq. 2). XY data are excluded from subsequent subfigures for clarity. **(b)** Out-of-plane magnetic field dependence of the SHG. **(c)** Temperature dependence of the SHG above T$_N$. Magenta insets indicate MnBi$_2$Te$_4$ layer number measured in each plot. All data in (b) and (c) are taken at a cryostat base temperature of ~4.5 K, and only XX data are shown for clarity.

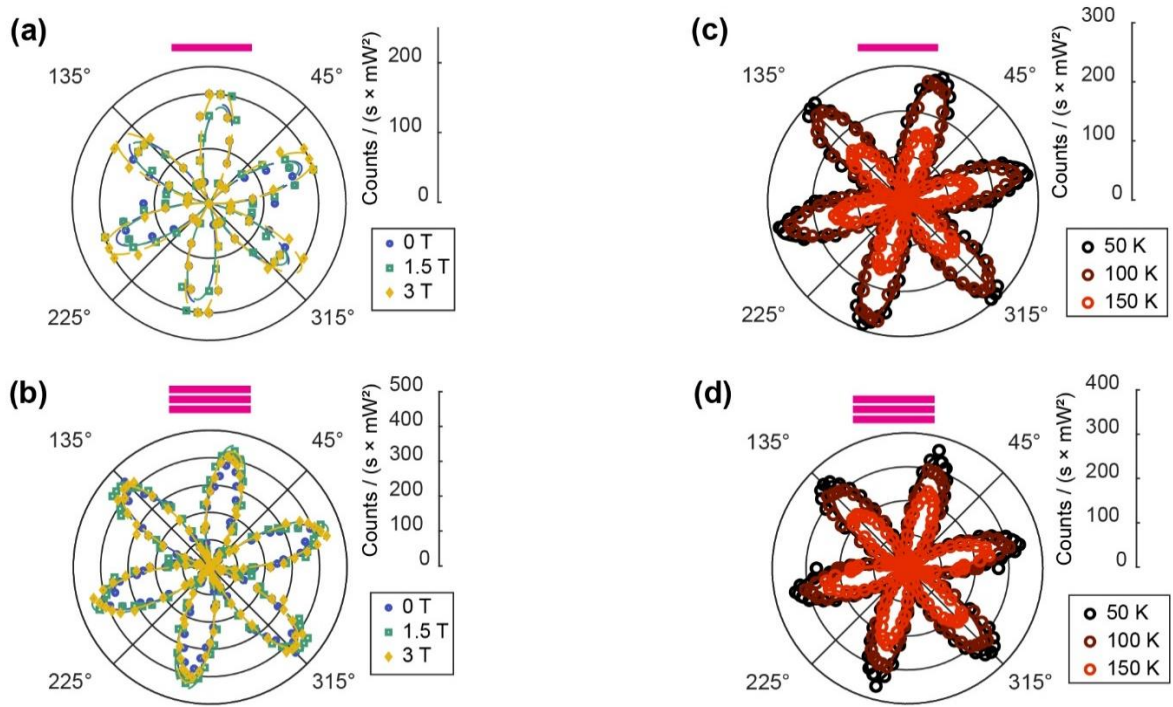

**Figure 3: 1 SL & 3 SL magneto- and temperature dependent SHG.** Dependence of the XX SHG on an out-of-plane magnetic field for 1 SL **(a)** and 3 SL flakes **(b)** at a base temperature of ~4.5 K. Temperature dependence of the XX SHG for 1 SL **(c)** and 3 SL **(d)** flakes above T$_N$. Magenta insets indicate MnBi$_2$Te$_4$ layer number. XY data, which exhibit the same behavior, are omitted for clarity.

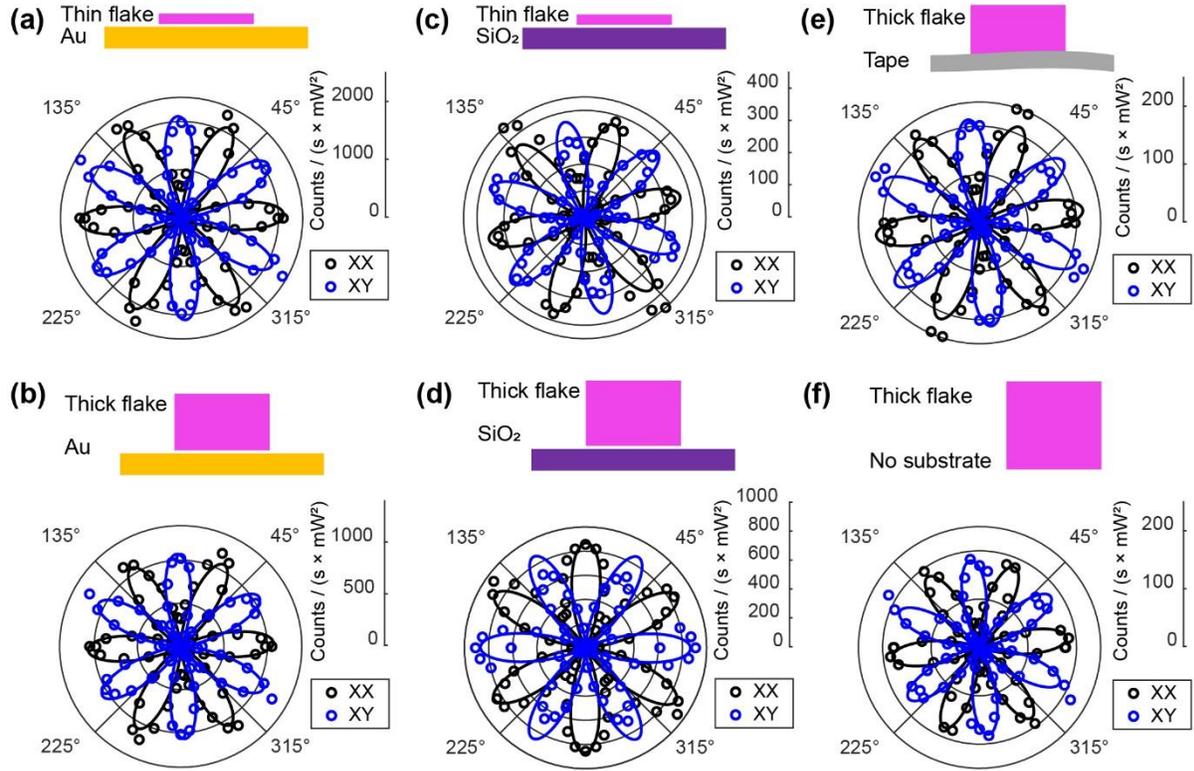

**Figure 4: Thickness and substrate dependence of polarization resolved SHG.** Polarization-resolved SHG from **(a)** a few-layer flake on Au, **(b)** a bulk flake on Au, **(c)** a few-layer flake on SiO$_2$, **(d)** Bulk flake on SiO$_2$, **(e)** a bulk flake cleaved on Scotch tape before exfoliation, and **(f)** a pristine bulk flake oriented to have the beam incident on the (0001) surface. All data are taken at cryogenic temperature with no applied magnetic field. Black (blue) dots are co(cross)-linearly polarized experimental data with solid line fits to Eq. 1 (Eq. 2).

For Table of Contents Only

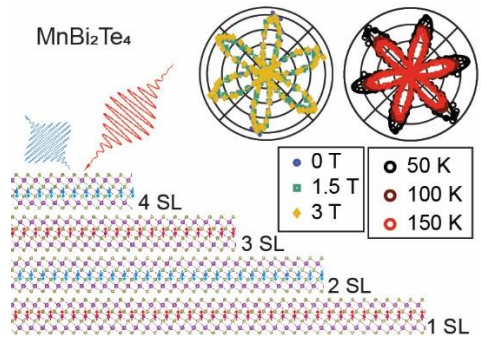

# Supplementary Information of

# Anomalous Second Harmonic Generation from Atomically Thin MnBi$_2$Te$_4$

**Authors:** Jordan Fonseca[1,†], Geoffrey M. Diederich[1,2,†], Dmitry Ovchinnikov[1], Jiaqi Cai[1], Chong Wang[3], Jiaqiang Yan[4,5], Di Xiao[1,3], Xiaodong Xu[1,3,*]

[1] Department of Physics, University of Washington, Seattle, Washington 98195, United States
[2] Intelligence Community Postdoctoral Research Fellowship Program, University of Washington, Seattle, Washington 98195, United States
[3] Department of Materials Science and Engineering, University of Washington, Seattle, Washington 98195, United States
[4] Materials Science and Technology Division, Oak Ridge National Laboratory, Oak Ridge, Tennessee 37831, United States
[5] Department of Materials Science and Engineering, University of Tennessee, Knoxville, Tennessee 37996, United States

[†] These authors contributed equally
[*] Correspondence to xuxd@uw.edu

**List of content:**

Methods

Figure S1: RMCD & SHG from a 4 SL sample

Figure S2: Comparison of SHG intensity from MnBi$_2$Te$_4$ and WSe$_2$

Figure S3: Effect of MnBi$_2$Te$_4$ surface degradation on SHG

Figure S4: Temperature and power dependent RMCD

# Methods

## Crystal growth and sample preparation

MnBi$_2$Te$_4$ bulk crystals were grown out of a Bi-Te flux as previously reported.[1] All samples were prepared inside an Argon-filled glovebox. We isolate thin flakes using standard scotch-tape exfoliation technique. To reliably procure flakes down to monolayer thickness, we evaporate a ~2 nm layer of Cr or Ti followed by a ~ 3 nm layer of Au onto 285 nm thick SiO$_2$/Si wafers and use the gold assisted exfoliation technique.[2,3] Optical and atomic force microscope images of few-layered flakes (Fig. 1c) were taken inside the glovebox before sealing flakes in a copper spacer and transferring them to a cold-finger 4 K cryostat for measurement. The schematic crystal structures in Fig. 1a and Fig. 1b were drawn using VESTA.[4]

## Optical measurements

All optical measurements were performed with light normally incident on the (0001) surface of the crystal. Measurements were conducted in a variable temperature cryostat including a superconducting magnet in Faraday geometry (Montana Instruments). The polar RMCD measurements were performed with a 632.8 nm HeNe laser, which was intensity-modulated by a mechanical chopper and polarization-modulated by a photoelastic modulator (Hinds PEM100). HeNe laser power was kept below 1 µW for all measurements. The SHG measurements were performed using femtosecond pulses from a 76 MHz Ti:sapphire oscillator (Coherent Mira 900F). The central output wavelength was chosen to be 800 nm for most measurements. The beam was focused onto the sample by a microscope objective (Olympus LUCPLFLN40X), which also collected the second harmonic signal. SHG is typically detected using a photomultiplier tube (Hamamatsu H8259) connected to a Time Interval and Frequency Counter (Stanford Research SR620). Polarization-resolved measurements were achieved by rotating the incident and collected beams together using a half-wave plate, while an additional half-wave plate and polarizer in the collection path allow for independent measurement of colinear (XX) and cross-linear (XY) SHG. The measurement is corrected for Faraday rotation in the objective and cryostat window by adjusting the rotation angles of the two waveplates based on calibrated measurements of the wavelength dependent Faraday rotation. To ensure that laser-induced heating does not melt magnetic ordering, we perform RMCD measurements at various ultrafast excitation powers and establish that at a cryostat base temperature of ~ 4.5K, laser powers below 500 µW do not locally heat the sample above T$_N$ (Fig. S4). To expedite measurements, all SHG data (except those shown in Fig. S2b) were collected for 180° of polarization. Data taken at angle $\theta$ were then duplicated to angle $\theta + 180°$. The validity of this approach was experimentally verified intermittently throughout our measurements.

## Rotational Anisotropy SHG Fitting

We compute the XX and XY polarization dependence of our measured SHG using the techniques described in.[5]

For the *2 SL AFM state*, the combination of P$_2$T symmetry and mirror symmetry along the $\Gamma - K$ line restricts the effective nonlinear tensor to

$$d_{eff} = \begin{vmatrix} d_{11} & -d_{11} & 0 & 0 & 0 & 0 \\ 0 & 0 & 0 & 0 & 0 & -d_{11} \\ 0 & 0 & 0 & 0 & 0 & 0 \end{vmatrix} \text{(ref. 6)}$$

With backscattering geometry, we can construct the electric fields of the excitation beam

$$E_x = E\cos(\theta - \varphi)$$
$$E_y = E\sin(\theta - \varphi)$$
$$E_z = 0$$

With $\theta$ the rotation angle of the light polarization and $\varphi$ an offset between the initial polarization and crystal axis. From this excitation beam, we calculate the nonlinear polarization induced in the crystal:

$$P_x = E^2 d_{11}(\cos(\theta - \varphi)^2 - E\sin(\theta - \varphi)^2)$$
$$P_y = -2E^2 d_{11}\sin(\theta - \varphi)\cos(\theta - \varphi)$$
$$P_z = 0$$

For comparison with our measurement, we convert these nonlinear polarizations into terms co- and cross-polarized with the normally incident beam using the relations

$$P_{xx} = P_x \cos(\theta - \varphi) + P_y \sin(\theta - \varphi)$$
$$P_{xy} = P_x \sin(\theta - \varphi) - P_y \cos(\theta - \varphi)$$

Finally, we model the shape of the normalized second harmonic response with

$$I_{xx}^{2\omega}(\theta, \varphi) \propto P_{xx}^2 = d_{11}^2 \cos^2[3(\theta - \varphi)] \quad \text{Equation 1}$$
$$I_{xy}^{2\omega}(\theta, \varphi) \propto P_{xy}^2 = d_{11}^2 \sin^2[3(\theta - \varphi)] \quad \text{Equation 2}$$

*Nonmagnetic* bulk MnBi$_2$Te$_4$ crystallizes in the space group $D_{3d}^{(5)}$, which is centrosymmetric with $d_{eff} = 0$ identically. At the surface, however, there is natural symmetry breaking at the interface between crystal and vacuum/substrate, which, like Bi$_2$Te$_3$ and Bi$_2$Se$_3$, has the point group symmetry $C_{3v}$,[7,8] consisting of a three-fold rotation about the $\hat{z}$ axis and three vertical mirror planes. This point group has the following effective nonlinear tensor

$$d_{eff} = \begin{vmatrix} 0 & 0 & 0 & 0 & d_{42} & -d_{22} \\ d_{22} & -d_{22} & 0 & d_{42} & 0 & 0 \\ d_{13} & d_{13} & d_{33} & 0 & 0 & 0 \end{vmatrix} \quad \text{(ref. 5)}$$

This form, despite its apparently greater complexity, simplifies for light normally incident on the (0001) surface to produce the same six-fold pattern, but with the position of XX and XY lobes reversed:

$$I_{xx}^{2\omega}(\theta, \varphi) \propto P_{xx}^2 = d_{22}^2 \sin^2[3(\theta - \varphi)]$$
$$I_{xy}^{2\omega}(\theta, \varphi) \propto P_{xy}^2 = d_{22}^2 \cos^2[3(\theta - \varphi)]$$

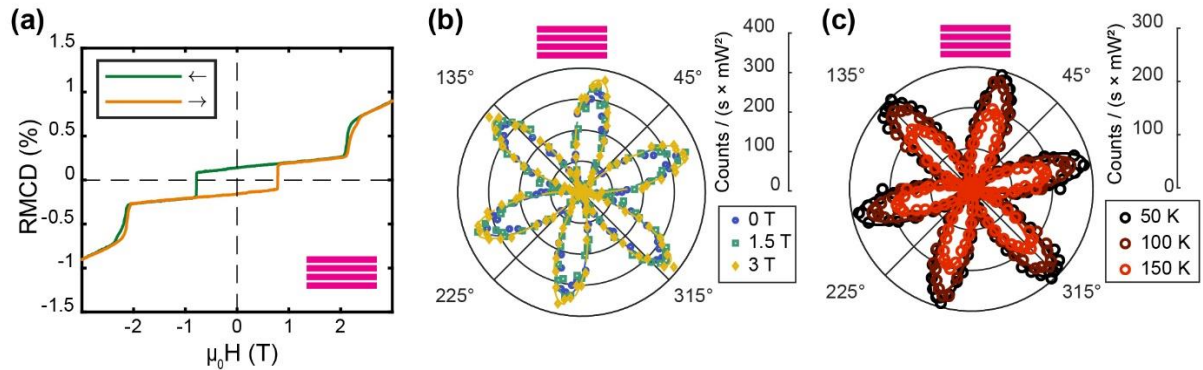

**Figure S1: RMCD & SHG from a 4 SL sample.** (a) RMCD from a 4 SL flake of MnBi$_2$Te$_4$ exfoliated on gold. (b) Magnetic field dependence of XX SHG from showing no change up to 3 T. (c) Temperature dependence of XX SHG flake showing a weak change in intensity of signal at higher temperatures, which we attribute to experimental artifacts.

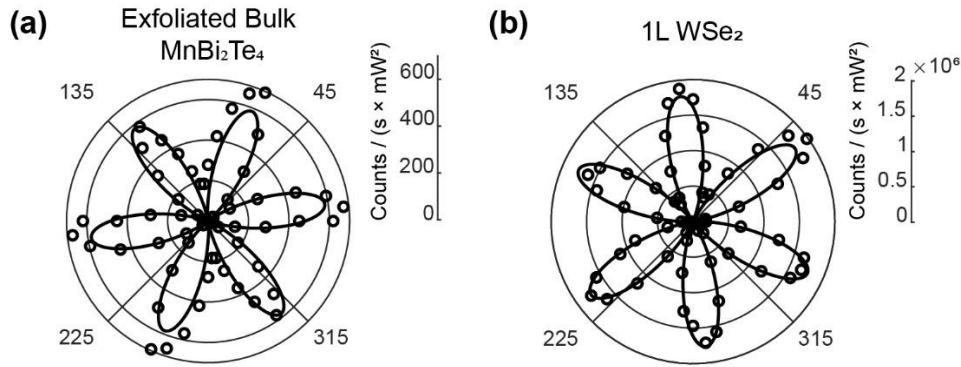

**Figure S2: Comparison of SHG intensity from MnBi$_2$Te$_4$ and WSe$_2$.** Polarization resolved SHG from (a) exfoliated bulk MnBi$_2$Te$_4$ and (b) 1L WSe$_2$ showing that the normalized intensity of SHG from MnBi$_2$Te$_4$ is approximately 3,000 times weaker than that from 1L WSe$_2$. All data are taken in the same experimental conditions other than excitation power, which was 850 µW in (a) and 600 µW in (b).

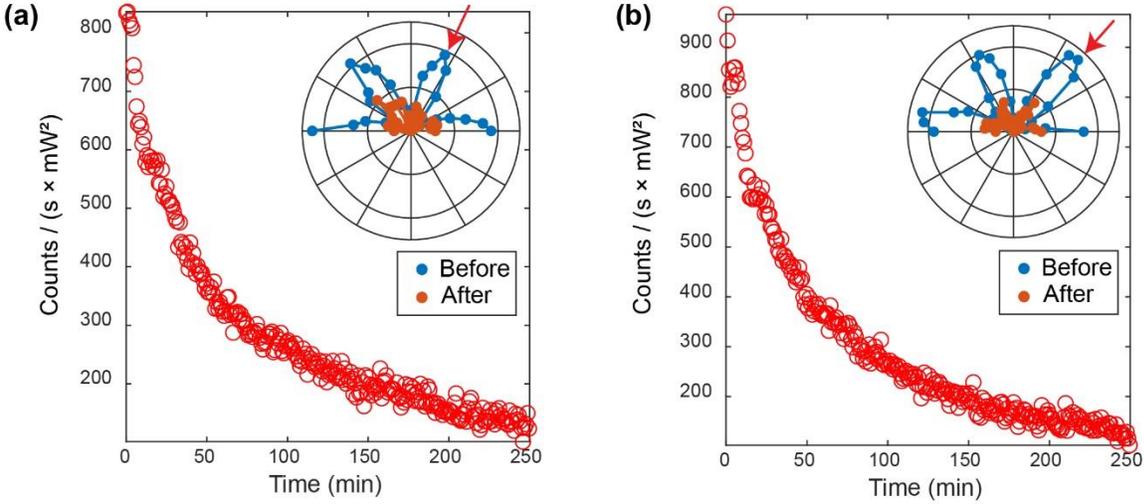

**Figure S3: Effect of MnBi$_2$Te$_4$ surface degradation on SHG**. Data shown in each panel taken from a different flake. Insets show a polarization resolved measurement before (blue) and after (orange) the time-dependent measurement. For the exposure time dependence, the SHG was collected at the polarization marked in the inset with a red arrow. SHG intensity decreases dramatically within the first 60 minutes following exfoliation.

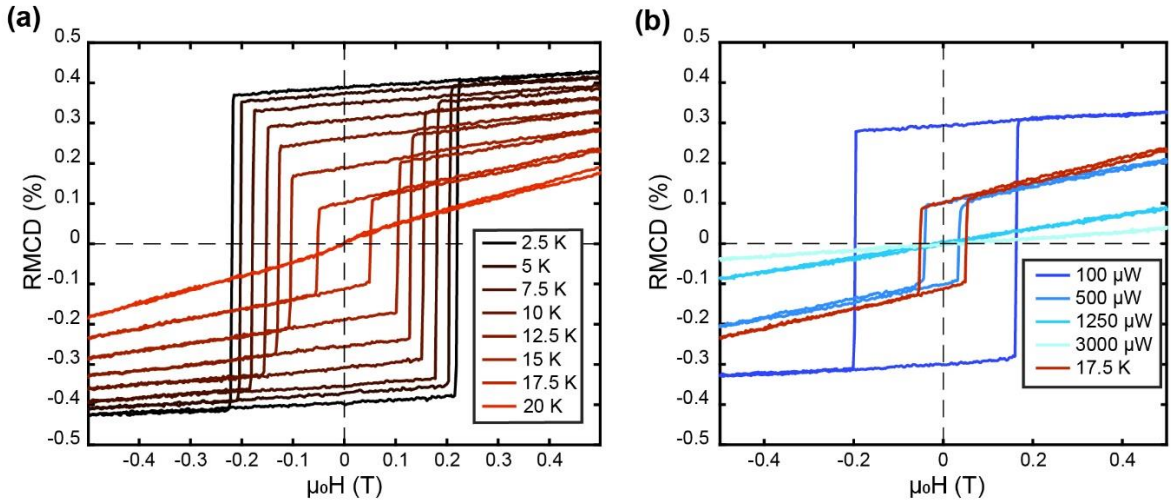

**Figure S4: Temperature and power dependent RMCD**. (a) Temperature dependent RMCD around the central hysteresis loop for a 3 SL sample. (b) RMCD signal for the same 3 SL measured as a function of ultrafast laser power when the SHG excitation beam is also on the sample. This data is overlapped with the RMCD data taken at 17.5 K when no ultrafast pulses were present. Below 500 µW, the local heating from the ultrafast beam does not raise the temperature of the 3 SL above its Néel temperature.